\documentclass[11pt,draft]{amsart}
\usepackage{amsmath}
\usepackage{amsfonts,amssymb,latexsym}
\usepackage{amscd}
\usepackage{mathrsfs}
\usepackage{amsthm}
\theoremstyle{plain}
\newtheorem{theorem}{Theorem} 
\newtheorem{corollary}{Corollary} 

\newtheorem{lemma}{Lemma} 
\newtheorem{proposition}{Proposition} 
 
\newtheorem{definition}{Definition}

\theoremstyle{remark}

\newtheorem{remark}{Remark}

\usepackage{graphics}
\numberwithin{equation}{section}

\def\logo{\raisebox{-10.5\p@}{\hb@xt@85\p@{\includegraphics{gft.eps}\hfil}}}

\def\un{1\kern-3pt \rm I}
\def\ptoday{{\ifcase\month 
\or January, \or February, \or March, \or April,\or May, 
\or June, \or July, \or August, \or September, \or October, 
\or November, \or December,\fi\ \number \year}}

\newcommand{\oN}{{\mathbb N}}
\newcommand{\oR}{{\mathbb R}}

\newcommand{\oC}{{\mathbb C}}

\input epsf
\usepackage[dvips]{graphicx}

\def\dj{\hbox{d\kern-0.347em \vrule width 0.3em height 1.252ex depth
-1.21ex \kern 0.051em}}

   \newcommand{\supp}{\mathrm{supp}}
   
\begin{document}

\title[A note on tempered ultrahyperfunctions]
      {\sl A Note on Fourier-Laplace Transform and \\[3mm]
       Analytic Wave front Set in Theory of \\[3mm] Tempered Ultrahyperfunctions}
        
\author{Daniel H.T. Franco}
\address{Centro de Estudos de F\'\i sica Te\'orica, Setor de F\'\i sica--Matem\'atica\\
         Rua Rio Grande do Norte 1053/302, Funcion\'arios \\
         Belo Horizonte, Minas Gerais, Brasil, CEP:30130-131.}
\email{dhtf@terra.com.br}

\author{Luiz H. Renoldi}
\address{Universidade Federal de S\~ao Jo\~ao del Rey\\
         Departamento de F\'\i sica, DCNAT\\  
         S\~ao Jo\~ao del Rei, Minas Gerais, Brasil, CEP:36300-000.}
\email{lhrenoldi@yahoo.com.br}

\date{January 30, 2006}
\thanks{L.H. Renoldi is supported by the Brazilian agency CAPES}

\begin{abstract}
In this paper we study the Fourier-Laplace transform of tempered ultrahyperfunctions
introduced by Sebasti\~ao e Silva and Hasumi. We establish a generalization
of Paley-Wiener-Schwartz theorem for this setting. This theorem
is interesting in connection with the microlocal analysis.
For this reason, the paper also contains a description
of the singularity structure of tempered ultrahyperfunctions
in terms of the concept of analytic wave front set.
\end{abstract}

\maketitle


\section{Introduction}
Tempered ultrahyperfunctions were introduced in papers of Sebasti\~ao e
Silva~\cite{Tiao1,Tiao2} and Hasumi~\cite{Hasumi}, under the name of tempered
ultradistributions, as the strong dual of the space of test functions
of rapidly decreasing entire functions in any horizontal strip.
While Sebasti\~ao e Silva~\cite{Tiao1} used extension procedures
for the Fourier transform combined with holomorphic representations
and considered the case of one variable, Hasumi~\cite{Hasumi} used
duality arguments in order to extend the notion
of tempered ultrahyperfunctions for the case of $n$-variables
(see also~\cite[Section 11]{Tiao2}). In a brief tour, Marimoto~\cite{Mari1}
gave some more precise informations concerning the work of Hasumi.
More recently, the relation between the tempered ultrahyperfuntions and
Schwartz distributions and some major results, as the kernel theorem and the
Fourier-Laplace transform have been established by Br\"uning and Nagamachi
in~\cite{BruNa1}. Further, aside from the mathematical interest of the
results presented in Refs.~\cite{Tiao1}-\cite{BruNa1}, Br\"uning and
Nagamachi have conjectured that the properties of tempered
ultrahyperfunctions are well adapted for their use in quantum field
theory with a fundamental length, while Bollini and Rocca~\cite{Bocca}
have given a general definition of convolution between two arbitrary tempered
ultrahyperfunctions in order to treat the problem of singular products
of functions Green also in quantum field theory. In another interesting
recent work~\cite{Andreas}, Schmidt has given an insight in the
operations of duality and Fourier transform on the space of test
and generalized functions belonging to new subclasses of Fourier
hyperfunctions of mixed type, satisfying polynomial growth conditions
at infinity, which is very similar to the studies by
Sebasti\~ao e Silva~\cite{Tiao1} and Hasumi~\cite{Hasumi} about
tempered ultrahyperfunctions, and eventually suggests applications
to quantum field theory.

In this article, we will give some precisions on the Fourier-Laplace
transform theorem for tempered ultrahyperfunctions, by considering
the theorem in its simplest form: the equivalence between support
properties of a distribution in a closed convex cone and the holomorphy
of its Fourier-Laplace transform in a suitable tube with conical basis.
All cones will have their vertices at the origin. After some preliminares
presented in Section \ref{SecUltra}, where for the sake of completeness we
include a brief exposition of the basic facts concerning the theory of tempered
ultrahyperfunctions, in Section \ref{SecTheo1} we define a space of functions
whose elements are holomorphic in tube domains corresponding to open convex cones.
In Section \ref{SecTheo2}, we extend the Paley-Wiener-Schwartz (PWS)
theorem for the setting of tempered ultrahyperfunctions by combining
two lemmas established in Section \ref{SecTheo1}. In this setting, the
PWS theorem deals with the Fourier-Laplace transform of 
distributions of exponential growth with support in a closed convex cone. This
result is also interesting in connection with the concept of analytic wave front
set. For this reason, in Section \ref{SecTheo3} we study the singularity structure
of tempered ultrahyperfunctions corresponding to an open cone $C \subset \oR^n$
via the notion of analytic wave front set, a refined description of the singularity
spectrum, with several applications all around Mathematics and Physics. Our aim is
to provide the microlocal analysis in the space of tempered ultrahyperfunctions
which is very similar to microlocal analysis in the framework of distributions. 

We note that the results obtained here are of importance in the construction
and study of nonstrictly localizable quantum field theories, namely, the
{\em quasilocal} field theories (where the fields are localizable only in
regions greater than a certain scale of nonlocality), and in fact they
have been motivated by recent results used in the axiomatic formulation of
quantum field theory with a minimum length~\cite{BruNa1}. The physical applications
of the results given in this paper will appear in a coming paper, in particular, to
quantum field theory in noncommutative spacetimes~\cite{DanHenri}.

\section{A Glance at the Theory of Tempered Ultrahyperfunctions:\\
Definitions and Basic Properties}
\label{SecUltra}
We shall recall in this paragraph some definitions and basic properties of the
tempered ultrahyperfunction space introduced by Sebasti\~ao e Silva~\cite{Tiao1,Tiao2}
and Hasumi~\cite{Hasumi}. We shall adopt here the point of view of not entering
into all technical aspects concerning the theory of tempered ultrahyperfunctions,
reminding the reader to consult the Refs.~\cite{Tiao1}-\cite{BruNa1} for more details.

\,\,\,{\bf Notations:} We will use the standard multi-index notation. Let
$\oR^n$ (resp. $\oC^n$) be the real (resp. complex) $n$-space whose generic points
are denoted by $x=(x_1,\ldots,x_n)$ (resp. $z=(z_1,\ldots,z_n)$), such that
$x+y=(x_1+y_1,\ldots,x_n+y_n)$, $\lambda x=(\lambda x_1,\ldots,\lambda x_n)$,
$x \geq 0$ means $x_1 \geq 0,\ldots,x_n \geq 0$, $\langle x,y \rangle=x_1y_1+\cdots+x_ny_n$
and $|x|=|x_1|+\cdots+|x_n|$. Moreover, we define
$\alpha=(\alpha_1,\ldots,\alpha_n) \in \oN^n_o$, where $\oN_o$ is the set
of non-negative integers, such that the length of $\alpha$ is the corresponding
$\ell^1$-norm $|\alpha|=\alpha_1+\cdots +\alpha_n$, $\alpha+\beta$ denotes
$(\alpha_1+\beta_1,\ldots,\alpha_n+\beta_n)$, $\alpha \geq \beta$ means
$(\alpha_1 \geq \beta_1,\ldots,\alpha_n \geq \beta_n)$, $\alpha!=
\alpha_1! \cdots \alpha_n!$, $x^\alpha=x_1^{\alpha_1}\ldots x_n^{\alpha_n}$,
and
\[
D^\alpha \varphi(x)=\frac{\partial^{|\alpha|}\varphi(x_1,\ldots,x_n)}
{\partial x_1^{\alpha_1}\partial x_2^{\alpha_1}\ldots\partial x_n^{\alpha_n}}\,\,.
\]

We consider two $n$-dimensional spaces -- $x$-space and $\xi$-space -- with the
Fourier transform defined
\[
\widehat{f}(\xi)={\mathscr F}[f(x)](\xi)=
\int_{\oR^n} f(x)e^{i \langle \xi,x \rangle} d^nx\,\,,
\]
while the Fourier inversion formula is
\[
f(x)={\mathscr F}^{-1}[\widehat{f}(\xi)](x)= \frac{1}{(2\pi)^n}
\int_{\oR^n} \widehat{f}(\xi)e^{-i \langle \xi,x \rangle} d^n\xi\,\,.
\]
The variable $\xi$ will always be taken real while $x$ will also be
complexified -- when it is complex, it will be noted $z=x+iy$.

\,\,\,We shall consider the function
\[
h_{K}(\xi)=\sup_{x \in K} \bigl| \langle \xi,x \rangle \bigr|\,\,,\quad \xi \in \oR^n\,\,,
\]
the indicator of $K$, where $K$ is a compact set in $\oR^n$. $h_{K}(\xi) < \infty$
for every $\xi \in \oR^n$ since $K$ is bounded. For sets
$K=\bigl[-k,k\bigr]^n$, $0 < k < \infty$, the indicator function $h_{K}(\xi)$ can
be easily determined:
\[
h_{K}(\xi)=\sup_{x \in K} \bigl| \langle \xi,x \rangle \bigr|=
k|\xi|\,\,,\quad \xi \in \oR^n\,\,,\quad |\xi|=\sum_{i=1}^n|\xi_i|\,\,.
\]
Let $K$ be a convex compact subset of $\oR^n$,
then $H_b(\oR^n;K)$ ($b$ stands for bounded) defines the space of all
functions $\in C^\infty(\oR^n)$ such that $e^{h_K(\xi)}D^\alpha f(\xi)$
is bounded in $\oR^n$ for any multi-index $\alpha$. One defines in
$H_b(\oR^n;K)$ seminorms
\begin{equation}
\|\varphi\|_{K,N}=\sup_{x \in \oR^n; \alpha \leq N}
\bigl\{e^{h_K(\xi)}|D^\alpha f(\xi)|\bigr\} < \infty\,\,,
\quad N=0,1,2,\ldots\,\,.
\label{snorma2}
\end{equation}

\begin{theorem}
The space $H_b(\oR^n;K)$ equipped with the topology given
by the seminorms (\ref{snorma2}) is a Fr\'echet space.
\end{theorem}

\begin{proof}
See~\cite{Hasumi,Mari1}. 
\end{proof}

If $K_1 \subset K_2$ are two compact convex sets, then
$h_{K_1}(\xi) \leq h_{K_2}(\xi)$, and thus the canonical
injection $H_b(\oR^n;K_2) \hookrightarrow H_b(\oR^n;K_1)$
is continuous. Let $O$ be a convex open set of $\oR^n$.
To define the topology of $H(\oR^n;O)$ it suffices to let $K$ range
over an increasing sequence of convex compact subsets $K_1,K_2,\ldots$
contained in $O$ such that for each $i=1,2,\ldots$,
$K_i \subset K_{i+1}^\circ$ ($K_{i+1}^\circ$ denotes the
interior of $K_{i+1}$) and ${O}=\bigcup_{i=1}^\infty K_i$.
Then the space $H(\oR^n;O)$ is the projective limit of the
spaces $H_b(\oR^n;K)$ according to restriction mappings
above, {\em i.e.}
\begin{equation}
H(\oR^n;O)=\underset{K \subset {O}}{\lim {\rm proj}}\,\,
H_b(\oR^n;K)\,\,,
\label{limproj2}
\end{equation} 
where $K$ runs through the convex compact sets contained in $O$.

\begin{theorem}
For the spaces introduced above the following statements hold:
\begin{enumerate}
\item The space ${\mathscr D}({\oR^n})$ of all $C^\infty$-functions
on $\oR^n$ with compact support is dense in $H(\oR^n;K)$ and $H(\oR^n;O)$.
\item The space $H(\oR^n;\oR^n)$ is dense in $H(\oR^n;O)$.
\end{enumerate}
\label{theoINJ}
\end{theorem}

\begin{proof}
See~\cite{Hasumi, Mari1}.
\end{proof}

From Theorem \ref{theoINJ} we have the following injections~\cite{Mari1}:
\[
H^\prime(\oR^n;K) \hookrightarrow H^\prime(\oR^n;\oR^n)
\hookrightarrow {\mathscr D}^\prime(\oR^n)\,\,,
\]
and
\[
H^\prime(\oR^n;O) \hookrightarrow H^\prime(\oR^n;\oR^n)
\hookrightarrow {\mathscr D}^\prime(\oR^n)\,\,.
\]

The dual space $H^\prime(\oR^n;O)$ of $H(\oR^n;O)$ is the space of
distributions $V$ of exponential growth~\cite{Mari1} such that
\[
V=D^\gamma_\xi[e^{h_K(\xi)}g(\xi)]\,\,,
\]
where $g(\xi)$ is a bounded continuous function.

Now, we pass to the definition of tempered ultrahyperfunctions.
In the space $\oC^n$ of $n$ complex variables $z_i=x_i+iy_i$,
$1 \leq i \leq n$, we denote by $T(\Omega)=\oR^n+i\Omega \subset \oC^n$
the tubular set of all points $z$, such that $y_i={\text{Im}}\,z_i$ belongs
to the domain $\Omega$, {\em i.e.}, $\Omega$ is a connected open set in $\oR^n$
called the basis of the tube $T(\Omega)$. Let $K$ be a convex compact
subset of $\oR^n$, then ${\mathfrak H}_b(T(K))$ defines
the space of all continuous functions $\varphi$ on $T(K)$ which are holomorphic
in the interior $T(K^\circ)$ of $T(K)$ such that the estimate
\begin{equation}
|\varphi(z)| \leq C (1+|z|)^{-N}
\label{est}
\end{equation}
is valid for some constant $C=C_{K,N}(\varphi)$. The best possible constants in
(\ref{est}) are given by a family of seminorms in ${\mathfrak H}_b(T(K))$
\begin{equation}
\|\varphi\|_{K,N}=\sup_{z \in T(K)}
\bigl\{(1+|z|)^{N}|\varphi(z)|\bigr\} < \infty\,\,,
\quad N=0,1,2,\ldots\,\,.
\label{snorma1}
\end{equation}

\begin{theorem}
The space ${\mathfrak H}_b(T(K))$ equipped with the topology given
by the seminorms (\ref{snorma1}) is a Fr\'echet space.
\end{theorem}

\begin{proof}
See~\cite{Hasumi,Mari1}. 
\end{proof}

The fact of the spaces ${\mathfrak H}_b(T(K))$ belong to the class of nuclear
Fr\'echet spaces is important for applications to QFT.

If $K_1 \subset K_2$ are two convex compact sets, then
${\mathfrak H}_b(T(K_2)) \hookrightarrow {\mathfrak H}_b(T(K_1)$.
Given that the spaces ${\mathfrak H}_b(T(K_i))$ are Fr\'echet spaces, the space
${\mathfrak H}(T({O}))$ is characterized as a projective limit of Fr\'echet spaces 
\begin{equation}
{\mathfrak H}(T({O}))=\underset{K \subset {O}}{\lim {\rm proj}}\,\,
{\mathfrak H}_b(T(K))\,\,,
\label{limproj1}
\end{equation}
where $K$ runs through the convex compact sets contained in $O$ and
the projective limit is taken following the restriction mappings above.

\begin{proposition}
If $f \in H(\oR^n;O)$, the Fourier transform of $f$ belongs
to the space ${\mathfrak H}(T(O))$, for any open convex
nonempty set $O \subset \oR^n$. By the dual Fourier transform
$H^\prime(\oR^n;O)$ is topologically isomorphic with the space
${\mathfrak H}^\prime(T(-O))$.
\label{Propo1}
\end{proposition}

\begin{proof}
See~\cite{Mari1}.
\end{proof}

\begin{definition}
A {\bf tempered ultrahyperfunction} is a continuous linear functional defined
on the space of test functions ${\mathfrak H}={\mathfrak H}(T(\oR^n))$ of rapidly
decreasing entire functions in any horizontal strip. The space of all tempered
ultrahyperfunctions is denoted by ${\mathscr U}(\oR^n)$.
\label{UHF}
\end{definition}

The space ${\mathscr U}(\oR^n)$ is characterized in the following way~\cite{Hasumi};
let $\boldsymbol{{\mathscr H}_\omega}$ be the space of all functions $f(z)$ such that:

\begin{itemize}

\item
$f(z)$ is analytic for $\{z \in \oC^n \mid |{\rm Im}\,z_1| > p,
|{\rm Im}\,z_2| > p,\ldots,|{\rm Im}\,z_n| > p\}$.

\item
$f(z)/z^p$ is bounded continuous  in
$\{z\in \oC^n \mid |{\rm Im}\,z_1| \geqq p,|{\rm Im}\,z_2| \geqq p,
\ldots,|{\rm Im}\,z_n| \geqq p\}$, where $p=0,1,2,\ldots$ depends on $f(z)$.

\item $f(z)$ is bounded by a power of $z$: $|f(z)|\leq C(1+|z|)^N$,
where $C$ and $N$ depend on $f(z)$.

\end{itemize}

Let $\boldsymbol{\Pi}$ be the set of all $z$-dependent pseudo-polynomials,
$z\in \oC^n$. Then ${\mathscr U}$ is the quotient space
${\mathscr U}=\boldsymbol{{\mathscr H}_\omega}/\boldsymbol{\Pi}$.
By a pseudo-polynomial we understand a function of $z$ of the form
$\sum_s z_j^s G(z_1,...,z_{j-1},z_{j+1},...,z_n)$, with
$G(z_1,...,z_{j-1},z_{j+1},...,z_n) \in \boldsymbol{{\mathscr H}_\omega}$.

According to Hasumi~\cite[Prop.5]{Hasumi} the dual ${\mathfrak H}^\prime$
of ${\mathfrak H}$ is algebraically isomorphic with the space ${\mathscr U}$.   

\section{Tempered Ultrahyperfunctions Corresponding \\to a Cone: 
The Space $\boldsymbol{{\mathscr H}^o_c}$}
\label{SecTheo1}
Let us introduce for the beginning some terminology and simple
facts concerning cones. An open set $C \subset \oR^n$ is called a cone if
$x \in C$ implies $\lambda x \in C$ for all $\lambda > 0$. Moreover,
$C$ is an open connected cone if $C$ is a cone and if $C$ is an open connected set.
In the sequel, it will be sufficient to assume for our purposes
that the open connected cone $C$ in $\oR^n$ is an open convex cone
with vertex at the origin. A cone $C^\prime$ is called compact in $C$ --
we write $C^\prime \Subset C$ -- if the projection ${\sf pr}{\overline C^{\,\prime}}
\overset{\text{def}}{=}{\overline C^{\,\prime}} \cap S^{n-1} \subset
{\sf pr}C\overset{\text{def}}{=}C \cap S^{n-1}$, where $S^{n-1}$ is the unit
sphere in $\oR^n$. Being given a cone $C$ in $x$-space, we associate with $C$ a
closed convex cone $C^*$ in $\xi$-space which is the set $C^*=\bigl\{\xi \in \oR^n
\mid \langle \xi,x \rangle \geq 0, \forall x \in C \bigr\}$.
The cone $C^*$ is called the {\em dual cone} of $C$ (see Fig. 1).
\begin{center}
\begin{picture}(100,50)(50,20)
\put(1.5,4){\line(5,-2){100}}
\multiput(1.5,4)(5,-2){21}{\line(-1,2){5}}
\put(101.5,-36){\line(5,2){100}}
\multiput(101.5,-36)(5,2){21}{\line(1,2){5}}
\put(95,-2){{\tiny${\sf pr}C$}}
\put(95,-13){{\tiny${\sf pr}C^\prime$}}
\put(83,32){{\tiny$C \setminus B[0;1]$}}
\put(193,11){{\tiny$C^*$}}
\put(101.5,-36){\line(2,5){40}}
\put(101.5,-36){\line(2,3){65}}
\put(157,57){{\tiny$C$}}
\put(130,57){{\tiny$C^\prime$}}
\put(101.5,-36){\line(-2,5){40}}
\put(101.5,-36){\line(-2,3){65}}
\qbezier(83.6,-9.5)(101.5,0)(119,-9.5)
\put(85,-50){\footnotesize{Figure 1}}
\end{picture}
\end{center}
\vspace{3cm}
By $T(C)$ we will denote the set $\oR^n+iC \subset \oC^n$. If $C$ is
open and connected, $T(C)$ is called the tubular radial domain in $\oC^n$,
while if $C$ is only open $T(C)$ is referred to as a tubular cone. An
important example of tubular radial domain in quantum field theory
is the forward light-cone
\[
V_+=\Bigl\{z \in \oC^n \mid {\rm Im}\,z_1 >
\Bigl(\sum_{i=2}^n {\rm Im}^2\,z_i \Bigr)^{\frac{1}{2}} \Bigr\}\,\,.  
\]

We will deal with tubes defined as the set of all points $z \in \oC^n$
such that
\[
T(C)=\Bigl\{x+iy \in \oC^n \mid
x \in \oR^n, y \in C, |y| < \delta \Bigr\}\,\,,
\]
where $\delta > 0$ is an arbitrary number. 

Let $C$ be an open convex cone and let $C^\prime$ be an arbitrary
compact cone of $C$. Let $B[0;r]$ denote a closed ball of the
origin in $\oR^n$ of radius $r$, where $r$ is an arbitrary positive
real number. Denote $T(C^\prime;r)=\oR^n+iC^\prime \setminus
\bigl(C^\prime \cap B[0;r]\bigr)$.
We are going to introduce a space of holomorphic functions
which satisfy certain estimate according to Carmichael and
Milton~\cite{Carmi2}. We want to consider the space
consisting of holomorphic functions $f(z)$ such that
\begin{equation}
\bigl|f(z)\bigr|\leq K(C^\prime)(1+|z|)^N e^{h_{C^*}(y)}\,\,,\quad
z=x+iy \in T(C^\prime;r)\,\,,
\label{eq31} 
\end{equation}
where $h_{C^*}(y)=\sup_{\xi \in C^*}|\langle \xi,y \rangle|$
is the indicator of $C^*$, $K(C^\prime)$ is a constant that depends
on an arbitrary compact cone $C^\prime$ and $N$ is a non-negative real number.
The set of all functions $f(z)$ which are holomorphic in $T(C^\prime;r)$ and
satisfy the estimate (\ref{eq31}) will be denoted by $\boldsymbol{{\mathscr H}^o_c}$.
In what follows, we shall prove two lemmas which will be important for our
extension of PWS theorem for the setting of tempered ultrahyperfunctions.
The proofs of lemmas are slight variations of that of Lemma 10 and Lemma
11 in~\cite{Carmi2}. Throughout the remainder of this paper $T(C^\prime;r)$ will
denote the set $\oR^n+iC^\prime \setminus \bigl(C^\prime \cap B[0;r]\bigr)$.

\begin{lemma}
Let $C$ be an open convex cone, and let $C^\prime$ be an arbitrary compact
cone contained in $C$. Let $h(\xi)=e^{k|\xi|}g(\xi)$, $\xi \in \oR^n$, be
a function with support in $C^*$, where $g(\xi)$ is a bounded continuous
function on $\oR^n$. Let $y$ be an arbitrary but fixed point of
$C^\prime \setminus \bigl(C^\prime \cap B[0;r]\bigr)$. Then
$e^{-\langle \xi,y \rangle}h(\xi) \in L^2$, as a function of $\xi \in \oR^n$.
\label{lemma0}
\end{lemma}

\begin{proof}
By Vladimirov~\cite[Lemma 2, p.223]{Vlad} there is a real number
$1 \geq c=c(C^\prime) > 0$ such that $\langle \xi,y \rangle \geq c |\xi||y|$
for every $\xi \in C^*$ and $y \in C^\prime$. Then, by using the fact that
$\sup_{\xi \in C^*} |g(\xi)| \leq M$, it follows that
\begin{equation}
\Bigl|e^{-\langle \xi,y \rangle}h(\xi)\Bigr| \leq
M e^{k|\xi|-c |\xi||y|}\,\,.
\label{eq31''}
\end{equation}
From (\ref{eq31''}) we have that
\begin{equation}
\int_{\oR^n}\Bigl|e^{-\langle \xi,y \rangle}h(\xi)\Bigr|^2 d\xi=
\int_{C^*}\Bigl|e^{-\langle \xi,y \rangle}h(\xi)\Bigr|^2 d\xi \leq
M^2 \int_{C^*}e^{-2(c |\xi||y|-k|\xi|)}d\xi\,\,.
\label{eq31p'}
\end{equation}
Using a result concerning the Lesbegue integral
(see Schwartz~\cite[Prop.32, p.39]{Sch}) and the assumption that $k < c|y|$
for fixed $k$, we get
\begin{equation}
\int_{\oR^n}\Bigl|e^{-\langle \xi,y \rangle}h(\xi)\Bigr|^2 d\xi
\leq M^2 S^{n-1} \int_0^\infty e^{-2(c|y|-k)t}t^{n-1}dt\,\,,
\label{eq31p''}
\end{equation}
where $S^{n-1}$ is the area of the unit sphere in $\oR^n$. Integrating
by parts $(n-1)$ times on the last integral in (\ref{eq31p''}), it
follows that
\begin{equation}
\int_{\oR^n}\Bigl|e^{-\langle \xi,y \rangle}h(\xi)\Bigr|^2 d\xi
\leq M^2 S^{n-1} (n-1)! (2c|y|-2k)^{-n}\,\,.
\label{eq31p'''}
\end{equation}
with $y$ fixed in $C^\prime \setminus \bigl(C^\prime \cap B[0;r]\bigr)$.
Thus the r.h.s. of (\ref{eq31p'''}) is finite. This implies that
$e^{-\langle \xi,y \rangle}h(\xi) \in L^2$, as a function of $\xi \in \oR^n$,
for $y$ fixed in $C^\prime \setminus \bigl(C^\prime \cap B[0;r]\bigr)$. 
\end{proof}

\begin{definition}
We denote by $H^\prime_{C^*}(\oR^n;O)$ the subspace of $H^\prime(\oR^n;O)$
of distributions of exponential growth with support in the cone $C^*$:
\begin{equation}
H^\prime_{C^*}(\oR^n;O)=\Bigl\{V \in H^\prime(\oR^n;O) \mid
\supp(V) \subseteq C^* \Bigr\}\,\,. 
\label{eq31'} 
\end{equation}
\end{definition}

\begin{lemma}
Let $C$ be an open convex cone, and let $C^\prime$ be an arbitrary compact
cone contained in $C$. Let $V=D^\gamma_\xi[e^{h_K(\xi)}g(\xi)]$, where
$g(\xi)$ is a bounded continuous function on $\oR^n$ and $h_K(\xi)=k|\xi|$
for a convex compact set $K=\bigl[-k,k\bigr]^n$. Let
$V \in H^\prime_{C^*}(\oR^n;O)$. Then $f(z)=(2\pi)^{-n}
(V,e^{-i\langle \xi,z \rangle})$ is an element of
$\boldsymbol{{\mathscr H}^o_c}$.
\label{lemma1}
\end{lemma}

\begin{proof}
The proof that $\supp(V) \subseteq C^*$ implies that $f(z)$ is holomorphic
in $T(C^\prime;r)$ is obtained by considering formula:
\begin{align}
f(z)=(2\pi)^{-n}(V,e^{-i\langle \xi,z \rangle})&=(2\pi)^{-n}\int_{C^*}
D^\gamma_\xi[e^{k|\xi|}g(\xi)] e^{-i\langle \xi,z \rangle} d^n\xi \nonumber \\
&=(2\pi)^{-n}(-i)^{|\gamma|} z^\gamma \int_{C^*} [e^{k|\xi|}g(\xi)]
e^{-i\langle \xi,z \rangle} d^n\xi\,\,.
\label{eq32}
\end{align}

In order to prove that $f(z)$ is holomorphic, it is enough to
consider the function
\begin{align}
h(z)=\int_{C^*} [e^{k|\xi|}g(\xi)] e^{-i\langle \xi,z \rangle} d^n\xi\,\,.
\label{eq33}
\end{align}
Let $z_o$ be an arbitrary but fixed point of $T(C^\prime;r)$ and let
$R(z_o;a) \subset T(C^\prime;r)$ be an arbitrary but fixed
neighborhood of $z_o$ with radius $a$, such that its closure is
in $T(C^\prime;r)$. Since $R(z_o;a)$ is fixed and has closure
in $T(C^\prime;r)$, we can find two balls of the origin in $\oR^n$
of radius $k$ and $\delta$, respectively, so that $0<r<k<|y|<\delta$
for all $y={\rm Im}(z)$, with $z=x+iy \in R(z_o;a)$ (see Fig. 2).
\begin{center}
\begin{picture}(100,50)(50,20)
\put(1.5,4){\line(5,-2){100}}
\multiput(1.5,4)(5,-2){21}{\line(-1,2){5}}
\put(101.5,-36){\line(5,2){100}}
\multiput(101.5,-36)(5,2){21}{\line(1,2){5}}
\put(101.5,24){\circle{100}}
\put(101.5,24){\vector(1,1){14}}
\put(100,22.8){{\tiny$\bullet$}}
\put(101.5,-36){\vector(1,4){7.5}}
\put(101.5,-36){\vector(-1,2){40}}
\put(101.5,-36){\vector(-1,4){4.8}}
\put(58,45){{\tiny$\delta$}}
\put(108,-4){{\tiny$k$}}
\put(95,-15){{\tiny$r$}}
\put(99,19.5){{\tiny$z_o$}}
\put(102,32){{\tiny$a$}}
\put(193,11){{\tiny$C^*$}}
\put(101.5,-36){\line(2,5){40}}
\put(101.5,-36){\line(2,3){65}}
\put(157,57){{\tiny$C$}}
\put(130,57){{\tiny$C^\prime$}}
\put(101.5,-36){\line(-2,5){40}}
\put(101.5,-36){\line(-2,3){65}}
\qbezier(90.2,-19.5)(101.5,-12.3)(112.2,-19.5)
\qbezier(83.6,-9.5)(101.5,0)(119,-9.5)
\qbezier(51.9,38)(101.5,72)(150.4,38)
\put(85,-50){\footnotesize{Figure 2}}
\end{picture}
\end{center}
\vspace{3cm}

Taking the absolute value of both sides of (\ref{eq33}) and using the
fact that $g(\xi)$ is bounded, we conclude that
\begin{align}
|h(z)|=\Bigl|\int_{C^*} [e^{k|\xi|}g(\xi)] e^{-i\langle \xi,z \rangle} d\xi \Bigr|
&\leq \int_{C^*} |g(\xi)| e^{k|\xi|+\langle \xi,y \rangle} d^n\xi \nonumber \\
&\leq \sup_{\xi \in C^*} |g(\xi)| \int_{C^*} e^{k|\xi|+\langle \xi,y \rangle} d^n\xi \nonumber \\
&\leq M \int_{C^*} e^{k|\xi|+\langle \xi,y \rangle} d^n\xi\,\,.
\label{eq34}
\end{align}
Choose an arbitrary but fixed $Y \in C^\prime$ such that $z=x+iY \in R(z_o;a)$.
Assume that $\xi$ belongs to the open half-space $\bigl\{\xi \in C^* \mid
\langle \xi,Y \rangle < 0\bigr\}$. Then, for some fixed number $c(Y) > 0$,
it follows that $\langle \xi,Y \rangle \leq -c(Y)|\xi|$ for $\xi \in C^*$.
Thus, with the assumption that $k < c(Y)$ for fixed $k$, we repeat part
of the argument used in proof of Lemma \ref{lemma0}, namely, we use
the result in Schwartz~\cite[Prop.32, p.39]{Sch} concerning the
Lesbegue integral to get
\begin{align}
|h(z)|\leq M S^{n-1}\!\!\int_0^\infty e^{-(c(Y)-k)t} t^{n-1}dt=
M S^{n-1} (n-1)! (c(Y)-k)^{-n}\,\,,
\label{eq35}
\end{align}
where $S^{n-1}$ is the area of the unit sphere in $\oR^n$.

Now, by differentiation of (\ref{eq33}), we immediately obtain that
\begin{align}
|D^\beta_z h(z)|&\leq M S^{n-1}\!\!\int_0^\infty
e^{-(c(Y)-k)t} t^{|\beta|+n-1}dt=
M S^{n-1} (n-1)! (c(Y)-k)^{-(|\beta|+n)}\,\,.
\label{eq36}
\end{align}
This shows that the integral defining $h(z)$ and any complex derivative,
$D^\beta_z h(z)$, converges uniformly for $z \in R(z_o;a)$.
Since $z_o$ is an arbitrary point in $T(C^\prime;r)$, it
follows that $h(z)$ exists and is holomorphic for $z \in T(C^\prime;r)$.
In turn, this implies that $f(z)$ exists and is holomorphic for $z \in T(C^\prime;r)$. 
From (\ref{eq32}) and (\ref{eq35}) it follows that the existence of a constant
$K(C^\prime)$ and a positive real number $N$ implies that
\[
|f(z)| \leq (2\pi)^{-n}|z^\gamma| |h(z)| \leq K(C^\prime)(1+|z|)^N\,\,\quad
z=x+iy \in T(C^\prime;r)\,\,. 
\]
Now, since for $y \in C^\prime \setminus \bigl(C^\prime \cap B[0;r]\bigr)$,
$\sup_{\xi \in C^*}e^{|\langle \xi,y \rangle|} > 1$, then
\[
|f(z)| \leq K(C^\prime)(1+|z|)^N \sup_{\xi \in C^*}e^{|\langle \xi,y \rangle|}
= K(C^\prime)(1+|z|)^N e^{h_{c^*}(y)}\,\,,
\]
for $z=x+iy \in T(C^\prime;r)$, from which follows the lemma. 
\end{proof}

\begin{remark}
A result as the Lemma \ref{lemma1} was obtained
by Carmichael and Milton~\cite{Carmi2} and Pathak~\cite{Pathak} to
other spaces of distributions. In~\cite{Carmi2} Carmichael
and Milton proved a result of this type for the dual spaces of the
spaces of type $\mathscr S$ introduced by Gel'fand and Shilov~\cite{Gel}.
Using techniques as in the paper of Carmichael and Milton~\cite{Carmi2},
Pathak~\cite{Pathak} proved similar result for tempered ultradistributions,
based on classes of ultradifferentiable functions.   
\end{remark} 

\section{A Generalization of the Paley-Wiener-Schwartz Theorem}
\label{SecTheo2}
In what follows, we shall show that more can be said concerning
the functions $f(z) \in \boldsymbol{{\mathscr H}^o_c}$. It will be shown that
$f(z) \in \boldsymbol{{\mathscr H}^o_c}$ can be recovered as the (inverse)
Fourier-Laplace transform\footnote{The convention of signs in the Fourier transform
which is used here one leads us to consider the inverse Fourier-Laplace
transform.} of the constructed distribution $V \in H^\prime_{C^*}(\oR^n;O)$.
This result is a generalization of the PWS theorem.

\begin{theorem}[Paley-Wiener-Schwartz-type Theorem]
Let $f(z) \in \boldsymbol{{\mathscr H}^o_c}$, where $C$ is an open convex cone.
Then the distribution $V \in H^\prime_{C^*}(\oR^n;O)$ has a uniquely
determined inverse Fourier-Laplace transform $f(z)=(2\pi)^{-n}
(V,e^{-i\langle \xi,z \rangle})$ which is holomorphic in
$T(C^\prime;r)$ and satisfies the estimate (\ref{eq31}).
\label{PWSTheo} 
\end{theorem}

\begin{proof}
Consider
\begin{equation}
h_y(\xi)=\int_{\oR^n}\frac{f(z)}{P(iz)}\,\,
e^{i\langle \xi,z \rangle}d^nx\,\,,\quad z \in T(C^\prime;r)\,\,, 
\label{eq41p}
\end{equation}
with $h_y(\xi)=e^{k|\xi|}g_y(\xi)$, where $g(\xi)$ is a bounded continuous
function on $\oR^n$, and $P(iz)=(-i)^{|\gamma|} z^\gamma$. By
hypothesis $f(z) \in \boldsymbol{{\mathscr H}^o_c}$ and satisfies (\ref{eq31}).
For this reason, for an $n$-tuple $\gamma=(\gamma_1,\ldots,\gamma_n)$
of non-negative integers conveniently chosen, we obtain
\begin{equation}
\Bigl|\frac{f(z)}{P(iz)}\Bigr|\leq
K(C^\prime)(1+|z|)^{-n-\varepsilon} e^{h_{c^*}(y)}\,\,,
\label{eq42} 
\end{equation}
where $n$ is the dimension and $\varepsilon$ is any fixed positive real
number. This implies that the function $h_y(\xi)$ exists and is a continuous function
of $\xi$. Further, by using arguments paralleling the analysis in~\cite[p.225]{Vlad}
and the Cauchy-Poincar\'e Theorem~\cite[p.198]{Vlad}, we can show that the
function $h_y(\xi)$ is independent of $y={\rm Im}\,z$. Therefore, we denote
the function $h_y(\xi)$ by $h(\xi)$.

From (\ref{eq42}) we have that $f(z)/P(iz) \in L^2$ as a function of
$x={\rm Re}\,z \in \oR^n$, $y \in C^\prime \setminus \bigl(C^\prime
\cap B[0;r]\bigr)$. Hence, from (\ref{eq41p}) and the Plancherel theorem
we have that $e^{-\langle \xi,y \rangle}h(\xi) \in L^2$ as a function of
$\xi \in \oR^n$, and  
\begin{equation}
\frac{f(z)}{P(iz)}={\mathscr F}^{-1}\bigl[e^{-\langle \xi,y \rangle}
h(\xi)\bigr](x)\,\,,\quad z \in T(C^\prime;r)\,\,,
\label{eq43} 
\end{equation}
where the inverse Fourier transform is in the $L^2$ sense. Here,
Parseval's equation holds:
\begin{equation}
(2\pi)^{-n}\int_{\oR^n}\Bigl|e^{-\langle \xi,y \rangle}h(\xi)\Bigr|^2d^n\xi=
\int_{\oR^n}\Bigl|\frac{f(z)}{P(iz)}\Bigr|^2
d^nx\,\,. 
\label{eq44}
\end{equation}
It should be noted that for Eq.(\ref{eq43}) to be true $\xi$ must belong to the
open half-space $\bigl\{\xi \in C^* \mid \langle \xi,y \rangle < 0\bigr\}$, for
$y \in C^\prime \setminus \bigl(C^\prime \cap B[0;r]\bigr)$, as stated by Lemma
\ref{lemma1}, since by hypothesis $f(z) \in \boldsymbol{{\mathscr H}^o_c}$.

Now, if $h(\xi) \in H^\prime_{C^*}(\oR^n;O)$, then $V=D^\gamma_\xi h(\xi)
\in H^\prime_{C^*}(\oR^n;O)$. Since $C^*$ is a regular set~\cite[pp.98, 99]{Sch},
thus $\supp(h)=\supp(V)$. By Lemma \ref{lemma1} $(V,e^{-i\langle \xi,z \rangle})$
exists as a holomorphic function of $z \in T(C^\prime;r)$ and satisfies the
estimate (\ref{eq31}). A simple calculation yields
\begin{equation}
(2\pi)^{-n}(V,e^{-i\langle \xi,z \rangle})=
P(iz){\mathscr F}^{-1}\bigl[e^{-\langle \xi,y \rangle}
h(\xi)\bigr](x)\,\,\quad z \in T(C^\prime;r)\,\,.
\label{eq45} 
\end{equation}
In view of Lemma \ref{lemma0}, the inverse Fourier transform can be interpreted
in $L^2$ sense. Combining (\ref{eq43}) and (\ref{eq45}), we have
$f(z)=(2\pi)^{-n}(V,e^{-i\langle \xi,z \rangle})$. The uniqueness follows
from the isomorphism of the dual Fourier transform, according to Proposition
\ref{Propo1}. This completes the proof of the theorem.  
\end{proof} 

The following corollary is immediate from Theorem \ref{PWSTheo}
and preceding construction:

\begin{corollary}
Let $C^*$ be a closed convex cone and $K$ a convex compact set in
$\oR^n$. Define an indicator function $h_{K,C^*}(y)$, $y \in \oR^n$,
and an open convex cone $C_K$ such that
$h_{K,C^*}(y)=\sup_{\xi \in C^*}\bigl|H_K(\xi)-\langle \xi,y \rangle\bigr|$
and $C_K=\bigl\{y \in \oR^n \mid h_{K,C^*}(y) < \infty\bigr\}$.
Then the distribution $V \in H^\prime_{C^*}(\oR^n;O)$ has a uniquely
determined inverse Fourier-Laplace transform $f(z)=(2\pi)^{-n}
(V,e^{-i\langle \xi,z \rangle})$ which is holomorphic in the tube
$T(C^\prime_K;r)=\oR^n+iC^\prime_K \setminus
\bigl(C^\prime_K \cap B[0;r]\bigr)$, and satisfies the following estimate,
for a suitable $K \subset O$,
\begin{equation}
\bigl|f(z)\bigr|\leq K(C^\prime)(1+|z|)^N e^{h_{K,C^*}(y)}\,\,,
\label{eq37} 
\end{equation}
where $C^\prime_K \Subset C_K$.
\end{corollary}

\begin{remark}
A result of this type has been established by Br\"uning and
Nagamachi~\cite[Thm.2.15]{BruNa1}. The space of holomorphic functions
$f(z)$ considered by Br\"uning and Nagamachi restricted to $C_K$
is a subspace of the space $\boldsymbol{{\mathscr H}^o_c}$ defined in this paper.
While the function $f(z)$ considered by Br\"uning and Nagamachi satisfies the growth
condition (\ref{eq37}) and is holomorphic in the interior
of $\oR^n \times iC_K$, in our case $f(z)$ satisfies (\ref{eq37})
but is required to be holomorphic in $\oR^n \times iC^\prime_K \setminus
\bigl(C^\prime_K \cap B[0;r]\bigr)$ only.   
\end{remark}

\section{Analytic Wave Front Set of Tempered Ultrapyperfunctions}
\label{SecTheo3}
This section is about the singularity structure of tempered ultrahyperfunctions.
Here, we shall follow the results and ideas of~\cite{Hor2} and characterize the
singularities of tempered ultrahyperfunctions via the notion of analytic wave front
set. Define ${\mathscr U}_c=\boldsymbol{{\mathscr H}^o_c}/\boldsymbol{\Pi}$ as
being the quotient space of $\boldsymbol{{\mathscr H}^o_c}$ by set of pseudo-polynomials.
The set ${\mathscr U}_c$ is the space of tempered ultrahyperfunctions corresponding
to the open cone $C \subset \oR^n$. Let us now consider the consequences of Theorem
\ref{PWSTheo}.

\begin{theorem}
If $u \in {\mathscr U}_c(\oR^n)$ and $V \in H^\prime_{C^*}(\oR^n;O)$, then
$WF_A(u) \subset \oR^n \times C^*$.
\end{theorem}

\begin{proof}[Indication of proof]
Taking into account that tempered ultrahyperfunctions are representable by means
of holomorphic functions, we use the integral representation of such objects
according to Proposition 11.1 in~\cite{Tiao1} (which has a similar
characterization for the case $n$-dimensional). Thus, according to Proposition
11.1 in~\cite{Tiao1}, every element $u \in {\mathscr U}_c(\oR^n)$ is representable
under the form
\begin{align}
u=\int_{\oR^n} V(\xi) e^{-i\langle \xi,z \rangle} d^n\xi=
\int_{C^*} V(\xi) e^{-i\langle \xi,z \rangle} d^n\xi=f(z)\,\,,
\label{eq41}
\end{align}
where $V$ is a distribution of exponential type.
Hence, we can determine the $WF_A(u)$ by just
looking at the behavior of $f(z)$, where $f(z)$ is any representative
of an element $u \in {\mathscr U}_c(\oR^n)$. By Paley-Wiener-Schawartz-type
theorem, Theorem \ref{PWSTheo}, $f(z)$ is holomorphic at $T(C^\prime;r)$
unless $\langle \xi,Y \rangle \geq 0$ for $\xi \in C^*$ and $Y \in C^\prime$,
with $|Y| < \delta$. Since $Y$ has an arbitrary direction in $C^\prime$,
this shows that
\[
WF_A(u) \subset \oR^n \times \bigl\{\xi \in \oR^n \setminus \{0\}
\mid \langle \xi,Y \rangle \geq 0 \bigr\}\,\,,
\]
which is the desired result. 
\end{proof}

\,\,\,{\bf Note Added:} After the text of the present paper was submitted for
publication, we learn that Carmichael already has published an article which
contains some similar results related to the our construction, especially
to the Sections \ref{SecTheo1} and \ref{SecTheo2}: R.D. Carmichael,
``{\em The tempered ultra-distributions of J. Sebasti\~ao e Silva,}''
{\bf Portugaliae Mathematica 36} (1977) 119. However, it should be
noted that the singularity structure of tempered ultrahyperfunctions,
here characterized by the analytic wave front set, has not been considered by
Carmichael. 


\end{document}